\documentstyle[emulateapj,psfig]{article}

\def\Mesz{M\'esz\'aros}
\def\Pacz{Paczy\'nski}
\def\De{De R\'ujula}
\def\beq{\begin{equation}}
\def\enq{\end{equation}}
\def\beqn{\begin{eqnarray}}
\def\enqn{\end{eqnarray}}
\def\eps{\varepsilon}

\begin{document}

\title{Cerenkov Line Emission as a Possible Mechanism of X-ray Lines in
Gamma-ray Bursts}
\author{W. Wang$^{1,*}$, Y. Zhao$^1$ and J. H. You$^2$}
\begin{center}
$^1$National Astronomical Observatories, Chinese Academy of Sciences, Beijing  100012, China \\
$^2$Department of Applied Physics, Shanghai Jiao Tong University, Shanghai 200030, China \\
$^*$E-mail:wwang@lamost.bao.ac.cn
\end{center}

\begin{abstract}
The recent discoveries of X-ray lines in the afterglows of
gamma-ray bursts (GRBs) provide significant clues to the nature of
GRB progenitors and central environments. However, the iron line
interpretation by fluorescence or recombination mechanism requires a 
large amount of iron material. We argue that
the very strong iron line could be attributed to an alternative
mechanism: Cerenkov line emission since relativistic electrons and 
dense medium exist near GRB sites. Therefore, the broad iron lines
are expected, and line intensity will be nearly independent of the
iron abundance, the medium with the anomalously high Fe abundance is
not required.
\end{abstract}

\keywords{gamma-rays: bursts - line: profiles - X-rays: general}

\section{Introduction}
Recent GRB observational reports have shown that the broad
emission line features during the X-ray afterglows of four GRBs
(GRB 970508, Piro et al. 1999; GRB 970828, Yoshida et al. 1999;
GRB 991216, Piro et al. 2000; GRB 000214, Antonelli et al. 2000).
X-ray lines of similar energies, are usually interpreted as iron
emission lines from fluorescence and recombination. According to the
variability and very large flux and equivalent width, we can
estimate that there are large iron masses of 0.01-1.0$M_\odot$ in
the emission region with a size of $\sim 10^{15}$cm. This much Fe
would have to come from the surrounding medium rather than GRB progenitors
to avoid the famous problem of Baryon contamination problem.

However, iron-rich medium may not be consistent with the normal
interstellar medium (ISM) observed, even in the favorable case of
dense regions of stellar formation. \Mesz\ \& Rees (1998) have
shown that the circumburst environment created by the stellar wind
before the hypernova (\Pacz\ 1998) could yield a line of
substantial intensity. And a similarly favorable situation is
expected in the supranova (Vietri \& Stella 1998), where a GRB is
preceded by a supernova explosion for several months to years with
ejection of an iron-rich massive shell. In these scenarios, high
density in excess of $10^{10}{\rm cm^{-3}}$ may be expected, and
Fe absorption features of strength comparable to those in emission
should also be detected. Alternative mechanisms for the production
of X-ray lines have also been proposed. Rees \& \Mesz\
(2000) have argued that the strong line emission can be attributed
to the interaction of a continuing post-burst relativistic outflow
from the central decaying magnetar with the progenitor stellar
envelope at distance less than a light hour, and only a small mass
of Fe is required. Another idea is provided in the frame that the
lines are not interpreted as iron lines but the strongly
Doppler-blueshift Ly$\alpha$ line emissions by jets of highly
relativistic cannonballs (Dar \& \De\ 2001).

In the present letter, we proposed a new mechanism for iron line
in X-ray afterglows: Cerenkov line emission. As we know, 
Cerenkov radiation will be 
produced when the particle velocity exceeds the light speed in the
medium. Furthermore, You \& Chen (1980) argued that for
relativistic electrons moving through a dense gas, the Cerenkov
effect will produce peculiar atomic/molecular emission lines:
Cerenkov lines. They have presented a series of formulae to
describe the properties of peculiar Cerenkov lines (You et al.
1984, 1986), and extended formulae can be applied to X-ray lines
such as Fe K edge lines (You et al. 2000, hereafter Y00). Elegant
experimental confirmation of Cerenkov lines in $O_2, Br_2$ gas and
$Na$ vapor using a $^{90}Sr\ \beta$-ray source has been obtained
in the laboratory (Xu et al. 1981, 1988, 1989). Cerenkov line
emission has following remarkable features: broad, asymmetrical,
Cerenkov redshift and polarized if relativistic electrons have an
anisotropic velocity distribution.

We emphasize the special importance of the Cerenkov redshift which
markedly strengthens the emergent intensity of the Cerenkov
emission line. For an optically
thick dense gas, the emergent line is determined both by the
emission and absorption. The absorption mechanisms for the normal
line and for the Cerenkov line are extremely
different. The intensity of normal line, $I^n$ is greatly weakened
by the strong resonant absorption $k_{lu}(\nu_{lu})$ (the
subscripts $u$ and $l$ denote the upper and lower-levels) due to
the fact that the normal line is located exactly at the position of
the intrinsic frequency $\nu_{lu}$ where $k_{lu}$ is very large.
In the extreme case of a very dense gas, the emergent flux has the
continuum with a black body spectrum, and the normal line
disappeared, $I^n \sim 0$. However, the Cerenkov line, located at
$\nu<\nu_{lu}$ due to the redshift, can avoid the resonant
absorption because of $k_{lu}(\nu<\nu_{lu})\to 0$. The main
absorption mechanism which affects the intensity of Cerenkov line
is the photoionization absorption $k_{bf}$ which is very small
compared with $k_{lu}$. Thus the Cerenkov line photons can easily
escape from deep inside a dense gas cloud, causing a strong
emergent line flux, only if the density of relativistic electrons
$N_e$ is high enough. In other words, the dense gas appears to be
more transparent for the Cerenkov line than normal line, which
makes it possible for the Cerenkov line mechanism to dominate over
the normal line mechanisms when the gas is very dense and there
exist abundant relativistic electrons in the emission region.

In the astrophysical processes of GRBs, a large amount of
relativistic electrons and dense gas regions could exist. For the
long burst afterglows localized so far, the host galaxies show
signs of star formation activity, where the high-density
environment is expected. Recent broadband observations of the
afterglows of GRB 000926 (Piro et al. 2001; Harrison et al. 2001)
and statistical analysis (Reichart \& Price 2002a, 2002b) imply
the evidence for a fireball in a dense medium. When the
ultra-relativistic blast waves interact with the dense regions,
very strong Cerenkov line emission in X-ray band would be
observed, and Cerenkov mechanism may dominate in the case of the
optically thick gases. The advantage of Cerenkov mechanism lies in
that Cerenkov line in the optically thick case is naturally broad and
the emergent intensity nearly independent of the abundance of
iron, as pointed out in the following section.

\section{Cerenkov Line Emission: Basic Formulae}
The essential point of the calculation of Cerenkov radiation is
the evaluation of the refraction index $n$ and extinction
coefficient $\kappa$. At a given frequency $\nu$, the larger the
index $n_\nu$, the larger the condition $v>c/n_\nu$ to be
satisfied and stronger the Cerenkov radiation at $\nu$ will be,
where $v$ is the particle velocity and $c$ is the light speed. We
adopt the simplified approximate formulae when we are only
concerned with the neighborhood of a given atomic line, $\nu \sim
\nu_{lu}$(Y00):
\beqn
n_\nu^2-1 &=& {c^3 \over
16\pi^3}\nu_{lu}^{-4}A_{ul}g_uN({S_l \over g_l}-{S_u \over g_u}), \\
\kappa_\nu &=& {c^3 \over 128\pi^4}\nu_{lu}^{-5}\Gamma_{lu}
A_{ul}g_uN({S_l \over g_l}-{S_u \over g_u}), \enqn where the
damping constant $\Gamma_{lu}$ is related to Einstein's
coefficient $A_{ul}$, $g$ is the statistical weights, $N$ the
number density of the concerned atoms, for example, for the
calculation of iron $K\alpha$, $N=N_{Fe}$ represents the density
of iron. $S_u$ and $S_l$ represent the actual occupation number of
electrons at levels $u$ and $l$ respectively. $y$ represents the
fractional energy displacement defined as $y\equiv -{\Delta\eps /
\eps_{lu}}$, where $\eps_{lu} \equiv h\nu_{lu}=\eps_u-\eps_l$, and
$y \ll 1$ because we are interested in the neighborhood of
$\nu_{lu}$.

It is known from the usual theory of Cerenkov radiation that the
spectral power $P_\nu$ emitted in unit frequency interval by an
electron moving with velocity $\beta=v/c$ is \beq P_\nu=4\pi({e^2
\over c})(1-\beta^{-2}n^{-2})\beta\nu. \enq Let $N(\gamma)d\gamma$
be the number of relativistic electrons per unit volume with
energies $\gamma=(1-\beta^{-2})^{-1/2}$ in the interval $d\gamma$,
assuming the lower and upper energy cut-offs $\gamma_1$ and
$\gamma_2$. Hence the emissivity in unit volume, unit solid angle
and unit frequency interval is \beq J_\nu={1 \over
4\pi}\int_{\gamma_1}^{\gamma_2}P_\nu N(\gamma)d\gamma \simeq {\pi
e^2 \over c}N_e\nu(n^2-1-\gamma_c^{-2})d\nu, \enq where $\gamma_c$
is the characteristic energy of electrons,
$\int_{\gamma_1}^{\gamma_2} N(\gamma)d\gamma / \gamma^2 \equiv
\gamma_c^{-2}N_e$. Replacing $\nu$ to $y$, $J_\nu^c d\nu=J_y^c
dy$, we have \beq J_y^c dy={\pi e^2 \over
c}N_e\nu^2(n^2-1-\gamma_c^{-2})dy. \enq Setting
$n^2-1-\gamma_c^{-2}=0$, we get the Cerenkov line width: \beqn
y_{lim} &\equiv& {\Delta\nu_{lim} \over \nu_{lu}}=C_0\gamma_c^2, \nonumber \\
C_0&=&{c^3 \over 16\pi^3}\nu_{lu}^{-4}A_{ul}g_uN({S_l \over
g_l}-{S_u \over g_u}). \enqn With Eqs. (1), (5) and (6), the
spectral emissivity becomes: \beqn
J_y^c dy&=&C_1N_e(y^{-1}-y_{lim}^{-1})dy, \nonumber \\
C_1&=&{e^2c^2 \over 16\pi^2}\nu_{lu}^{-2}A_{ul}g_uN({S_l \over g_l}-{S_u \over
g_u}).
\enqn

In an optically thick dense gas, the Cerenkov line mechanism may be efficient,
the final intensity $I^c$ is determined by the competition between the
emission $J^c$ and absorption $k$. The main absorption mechanisms which affect
the emergent intensity of Cerenkov line are the line absorption $k_{lu}$ and
photoionization absorption $k_{bf}$ (free-free absorption can be neglected in
X-ray band), then the total absorption coefficient is $k=k_{lu}+k_{bf}$.
$k_{lu}$ can be easily obtained, $k_{lu}=4\pi\nu_{lu}\kappa_\nu /c$, using Eq.
(2), we get:
\beqn
k_{lu}&=&C_2y^{-2}, \nonumber \\
C_2&=&{c \over 32\pi^3}\nu_{lu}^{-4}\Gamma_{lu} A_{ul}g_uN({S_l
\over g_l}-{S_u \over g_u}).
\enqn

The photoionization absorption coefficient is
$k_{bf}=\sum\limits_{s}N_s\sigma_{bf}(s)$, where $\sigma_{bf}(s)$
is the photoelectric cross section of level $s$. If we investigate
it in Fe K edge lines, the dominate absorber should be the iron
atoms or ions rather than other elements. Thus, for Fe $K\alpha$
line: \beq k_{bf}\simeq
N_{Fe}S_2\sigma_{bf}(2)=8.4\times10^{-46}N_{Fe}S_2\eps_{lu}^{-3},
\enq where $\eps_{lu}=\eps_{21}=6.4$ keV. Only $s=2$ energy level
is considered due to the fact that around 6.4 keV,
$\sigma_{bf}(2)\gg\sigma_{bf}(s>2)$.

Finally, we can estimate the Cerenkov line radiation intensity in
the optically thick case: \beq I_y^c \simeq {J_y^c \over k}={J_y^c
\over k_{lu}+k_{bf}} = {N_eC_1(y^{-1}-y_{lim}^{-1}) \over
C_2y^{-2}+k_{bf}}. \enq  From the equation above, we see that both
the emissivity and absorption are proportional to the density of
iron atoms $N_{Fe}$, then the emergent intensity is only weakly
dependent on $N_{Fe}$ due to the existence of factor
$y_{lim}^{-1}$ in expression. The property of independence on iron
density could be used to resolve the puzzle of excessive amount of
iron in the ISM. Integrating Eq. (10), the total intensity of
Cerenkov line is obtained as \beq I^c=\int_{0}^{y_{lim}}I_y^c
dy=Y[ln(1+X^2)+2(1-{arctanX \over X})], \enq where $Y\equiv {N_e
\over 2}{C_1 \over k_{bf}}$, and $X\equiv \sqrt{k_{bf} \over
C_2}y_{lim}$.

\section{X-ray Lines in GRBs}
We believe the existence of Cerenkov line mechanism operating in
GRBs because there definitely exist dense gas regions and abundant
relativistic electrons, both of which are just the right
conditions to produce the Cerenkov line emission. In this section,
we will discuss the iron lines by Cerenkov line mechanism in the
X-ray afterglows of GRBs. We relist the parameters $C_0, C_1, C_2$
and $k_{bf}$ in X-ray band (in the normal conditions of a gaseous
medium, we have the approximation $({S_l \over g_l}-{S_u \over
g_u})\simeq {S_l \over g_l})$:
\beqn
C_0&=&1.05\times10^{-76}\eps_{lu}^{-4}A_{ul}g_uN_{Fe}{S_l \over
g_l}, \nonumber \\
C_1&=&5.77\times10^{-53}\eps_{lu}^{-2}A_{ul}g_uN_{Fe}{S_l \over
g_l}, \nonumber \\
C_2&=&1.75\times10^{-87}\eps_{lu}^{-4}\Gamma_{lu}A_{ul}g_uN_{Fe}{S_l \over
g_l}, \nonumber \\
k_{bf}&=&8.4\times10^{-46}\eps_{lu}^{-3}N_{Fe}S_2.
\enqn

In the case of the Fe $K\alpha$ line, only levels $l=1$ and $u=2$ are
considered. Because of the high electron temperature $T\sim10^7$K
in the condition of GRB afterglows, $S_1=g_1=2, g_2=8$, and
$S_2\sim5$. In the optically thick gas, we have found $X\gg1$,
then we can describe the formulae of Cerenkov line intensity as
$I^c\simeq 2Y(lnX-1)$. And we only take the main contribution $2Y$
in our following estimation, so \beq I^c\sim{C_1\over
k_{bf}}N_e\sim1.12\times10^{-7}\eps_{12}A_{21}N_e. \enq For the
optically thick case, the outward flux per unit area and time $\pi
F \approx \pi I^c$. Assuming a lot of spherical clouds with dense
gas that is distributed homogeneously in the circumburst environment,
the cloud radius and number are $R, N_c$ respectively, an isotropic
fireball interacts with these clouds. Therefore, the total line
luminosity from the clouds is $L^c=4\pi^2 R^2N_cI^c$. Defining a
covering factor $C \equiv {N_c\pi R^2/{4\pi D^2}}$, where $D$ is
the distance between the clouds and burst center. Then \beq
L^c=16\pi^2CD^2I^c\sim 8.2\times10^{32}C_{0.1}D_{16}^2N_e{\rm erg\
s^{-1}}, \enq where we have taken $\eps_{12}=6.4{\rm keV},
A_{21}=4.6\times10^{14}{\rm s^{-1}}$ and the characteristic scales
$C\sim 0.1, D \sim 10^{16}$cm.\footnote{Since Cerenkov emission is
from the gaseous medium, not from the relativistic electrons, so
the blue-shifted effect is not prominent.} From Eq. (14), we can
clearly see that the total line luminosity by the Cerenkov mechanism
is strongly dependent on the electron density rather than the iron
abundance. Thus, we propose that some classes of GRB models predict
the presence of a very high relativistic electron density, therefore the
ultra-strong iron lines in the X-ray afterglows can be explained
probably without additional request for initial and external
conditions of iron-rich torus.

We also could estimate the relativistic electron density required
for X-ray lines from four GRB afterglows. In the computations, we
assume that the X-ray lines in GRB 970828 and 000214 are also Fe
$K\alpha$ lines, and the line with a central energy of 3.49 keV in
GRB 991216 is considered (the other in 4.4 keV is explained as the
recombination continuum of H-like iron in 9.28 keV). Take the
standard cosmological model with $H_0=70{\rm km\ s^{-1}\
Mpc^{-1}}, \Omega_0=1, \Lambda=0$, and let $C_{0.1}=D_{16}=1$, our
final results and GRB line parameters are displayed in Table 1. We
noticed the electron densities $N_e\sim 10^{10}-10^{11}{\rm
cm^{-3}}$ are similiar to those estimated for the delayed
energy injection (Rees \& \Mesz\ 2000) and supranova model
(Vietri et al. 2001). We first simply estimate the relativistic
electron density from the fireball, $N_e \sim E/{4\pi D^2\Delta_D
\gamma m_pc^2}$, and taking the isotropic energy $E\sim 10^{53}$
erg, $\gamma \sim 100$, $\Delta_D \sim \gamma cT$ (T is the
duration $\sim 10$s), we obtained the density $N_e\sim 10^8 {\rm
cm^{-3}}$, which will be lower than our requirement in the model.
So we think the relativistic electrons should come from other
processes. We here propose that the electrons would be produced by 
a pulsar wind from the central millisecond magnetar (Thompson 1994)
or $e^+ e^-$ outflow from the Kerr black hole with magnetized
torus (MacFadyen \& Woosley 1999) as the delayed injection after
GRB events. Our interpretation is similar to that of Rees \&
\Mesz\ (2000), but we consider the $e^+ e^-$ outflow instead of
electromagnetic flux. The luminosity is required as high as $L\sim
4\pi D^2N_e c\gamma_e m_e c^2 \sim 10^{48} {\rm erg\ s^{-1}}$,
where we take $N_e \sim 10^{10}{\rm cm^{-3}}, \gamma_e \sim 10$ at 
the distance of $10^{16}$cm. The luminosity could be satisfied
by the central compact objects with the acceptable parameters such
as very strong magnetic field.

Because the scattering cross section of relativistic electrons
will be near to the zero due to the Klein-Nishina formula in the
very high frequency(due to the Doppler effect), we need not worry
about the electron scattering process will greatly effect the
line profile and intensity. The line width is also comparable to
the observed one in GRBs, which has been displayed in Fig. 5 of
the previous work(Y00).

\section{Discussion}
The recently detected X-ray lines in the afterglows of GRBs give us important
information on the nature of GRB progenitors by imposing significant
constraints on models and the severe problem may be how to arrange a huge
amount of iron-rich material close to GRB sites, while avoiding a very large 
opacity in the same time. In this Letter, we have presented an alternative
model of the X-ray spectral features by Cerenkov line mechanism, which would
not have the above problems.

The broad feature of Cerenkov line emissions is consistent with
the observations. Due to the Cerenkov redshift, the Cerenkov mechanism can
avoid strong absorption, and may dominate line mechanism in the
optically thick medium. What's more, in the previous section, we
have estimated the Cerenkov line luminosity in GRBs. We find that
it is nearly independent of iron abundance, a large amount of Fe is not 
required as the model of Rees \& \Mesz\ (2000), so additional physical
mechanisms involving the iron-rich medium before triggering GRBs
are not needed under our interpretation. And the line intensity
can be comparable to the observed value with the relativistic
electron density around $10^{10}-10^{11}{\rm cm^{-3}}$, which
could be satisfied in the environment of GRBs as the delayed
injection from central objects. Because only a simple analysis is
presented in this letter, detailed numerical calculations in all
cases of media will be developed in our following work.

Furthermore, we present a few arguments for fluorescence and
recombination mechanisms. The fluorescence model predicts positive
correlation of both line curves and flux between the $K\alpha$
line and X-ray continuum, which need further observations to
confirm. Besides, the prediction of marked absorption dip at edge
$>7$ keV which always accompanies the fluorescent $K\alpha$ line
(Young et al. 1998) has not been observed. The recombination
mechanism also involves serious problems. Recombination lines
require a very high temperature, which should be considered by
further theoretical models. Because of a very large optical depth,
the ionization edge at 9.28 keV would appear as an absorption
rather than an emission feature (Rees \& \Mesz\ 2000). Up to now,
a discovery of a transient absorption edge in the X-ray spectrum of
GRB 990705 (Amati et al. 2000) has been reported, however, it is
still surprising that no absorption features are observed in X-ray
afterglows in such a dense medium. The problems need more detailed
investigations.

We have shown that fluorescence and recombination are not the only
possible mechanisms for producing the observed lines in X-ray afterglows 
of GRBs. Then a question is put forward: how to discriminate the Cerenkov
line mechanism with them? Cerenkov lines have some interesting
features: broad, asymmetry, Cerenkov redshift and polarization
with the anisotropic velocity distribution of electrons.
Therefore, we expect that future observations of high spectral
resolution missions such as {\it Chandra}, XMM-Newton and future
mission {\it Swift} could address these important issues,
providing us more information of GRB progenitors and circumburst
environment.

We are grateful to Tan Lu, Zigao Dai, Xiangyu Wang and Zhuo Li for
the fruitful discussions.

\begin{table}
\caption{The GRB line parameters and derived electron densities}
\begin{center}
\begin{tabular}{l c c c c c c}
\tableline \tableline GRB & $z$ & $E$ & $\Delta E$ & $I_{line}$ &
$L_{line}$ & $N_e({\rm cm^{-3}})$
\\
\tableline 970508 & 0.835 & 3.4 & $\leq0.5$ & $5.0\pm2.0$ &
$2.7\times10^{44}$ &
$3.0\times10^{11}$ \\
970828 & 0.957 & 5.04 & $0.31_{-0.31}^{+0.38}$ & $1.9\pm1.0$ &
$1.1\times10^{44}$ & $1.5\times10^{11}$ \\
991216 & 1.02 & 3.49 & $0.23\pm0.07$ & $3.2\pm0.8$ &
$4.0\times10^{44}$ &
$4.8\times10^{11}$ \\
000214 & 0.47 & 4.7 & 0.2 & $0.9\pm0.3$ & $4.0\times10^{43}$ &
$4.8\times10^{10}$ \\
\end{tabular}
\end{center}
\tablecomments{$z$ is the cosmological redshift, the redshift of
GRB 000214 corresponds to Fe $K\alpha$ emission interpretation.
$E$ is the centroid energy of lines with width $\Delta E$ in keV
units. $I_{line}$ is the observed line intensity in units of
$10^{-5}{\rm photon\ cm^{-2}\ s^{-1}}$, and $L_{line}$ is the
luminosity in erg\ s$^{-1}$ units. See the details in the text.}
\end{table}

\end{document}